\documentclass[reprint,aps,showpacs,floatfix,10pt,twocolumn]{revtex4-1}
\pdfoutput=1
\usepackage{graphicx}
\usepackage{amssymb}
\usepackage{graphics, color}
\usepackage{amsmath}
\usepackage{rotating}
\usepackage{caption}  
\usepackage{setspace}  
\usepackage[font={stretch=1.2, footnotesize},labelfont=bf,
   justification=justified,
   format=plain, 
   singlelinecheck=false]{caption} 
\usepackage{lipsum} 
\usepackage{overpic}
\usepackage{float}
\usepackage[english]{babel}
\usepackage{hyperref}
\usepackage[usenames,dvipsnames]{xcolor}
\usepackage{adjustbox}
\usepackage{amsmath,amsfonts,amssymb,verbatim,float,mathtools}
\usepackage[titletoc,title]{appendix} 
\usepackage{caption}
\newcommand{\Lim}[1]{\raisebox{0.5ex}{\scalebox{0.8}{$\displaystyle \lim_{#1}\;$}}}
\usepackage{graphicx}
\usepackage{amsmath}
\graphicspath{{pics/}}
\hypersetup{%
	pdfborder = {0 0 0}}

\begin{document}
\title{Stable Collective Dynamics of Two-Level Systems Coupled by Dipole Interactions}
\author{C. D. Parmee}
\affiliation{T.C.M. Group, Cavendish Laboratory, University of Cambridge, JJ Thomson Avenue, Cambridge, CB3 0HE, U.K.}
\author{N. R. Cooper}
\affiliation{T.C.M. Group, Cavendish Laboratory, University of Cambridge, JJ Thomson Avenue, Cambridge, CB3 0HE, U.K.}
\begin{abstract}
We study the dynamics of a set of two-level systems coupled by dipolar interactions under a resonant external Rabi drive. The two-level systems are  prepared initially in a coherent product state, and we ask how the non-equilibrium conditions caused by the drive affect this coherence.
We study the full non-linear dynamics of the coupled two-level systems within a classical approximation by analysing numerically the equations of motion and determining the stability of the collective coherent state within classical Floquet theory. We establish the behaviour analytically in the high Rabi coupling limit by employing a Magnus expansion and spin wave analysis. Our results show that, typically, the dipole interactions between the two-level systems lead to instabilities that cause a breakdown of the collective Rabi oscillations. However, we identify parameter regimes for which the two-level systems undergo collective coherent Rabi oscillations even in the presence of the dipole interactions.
\end{abstract}
\date{\today} 
 \maketitle


\section{Introduction}

Systems consisting of a macroscopic number of degrees of freedom can present interesting forms of collective behaviour, notably the existence of different phases of matter at thermal equilibrium. Much interest is currently focussed on macroscopic quantum systems in far-from-equilibrium settings. While it could be expected that interactions will lead to a redistribution of energy among all degrees of freedom, leading to thermalization at long times, it is known that this need not always occur. Instead, there can be regimes of synchronization, many-body localization or prethermalization \cite{Eisert2015,Reimann2016,Nandkishore2015,Langen2014} that can take place instead. Determining under what conditions a system does not thermalize is an active area of current research.

One of the simplest physical realisations which explores the far-from-equilibrium dynamics of many interacting quantum degrees of freedom is that of Rabi oscillations in coupled two-level systems. The coupling arises from interactions, such as dipole interactions if the two-level systems carry dipole moments. Such settings occur naturally in nuclear spin systems, but can also be engineered in cold atom gases. With careful control over the strength of the interactions and geometry, cold gases allow the study of the long-time dynamics of a well-isolated system. Indeed, experiments of dipole-coupled two-level systems have already been performed in cold gases using Rydberg atoms \cite{Maxwell2013,Maxwell2014,Barredo2015,Bettelli2013} or polar molecules \cite{Gorshkov2011,Barnett2006,Gorshkov2011a,Micheli2006,Schachenmayer2011}. The interactions between the two-level systems leads to (local) energy shifts of the resonance frequency. Consequently, generically, when driven out of equilibrium the interactions are expected to lead to dephasing of collective Rabi oscillation \cite{Ota2007,Olmos2010,Jin2013,Hazzard2014,Fedorov2006,Ates2012,Lesanovsky2010} causing observables to relax to some (non-equilibrium) steady state values

In this paper, we show that dipolar interactions need not lead to dephasing of the collective Rabi oscillation at the mean-field level.
We consider a closed system of a large number of two-level systems which are coupled by dipole interactions and subjected to an external drive. After preparing the system out of equilibrium, we seek to find when collective coherent behaviour occurs, as a function of the strength of the external drive and also dependence on the form of the dipole-coupling (modified by the geometry, by the orientation of the dipoles, and by the relative strength of the static and transition dipole moments). Whereas the interactions give rise to chaotic behaviour and possible thermalisation in many cases, we show that there are parameter regimes where this no longer occurs and the system undergoes stable collective Rabi oscillations.
By showing that stable collective oscillations can occur under an external drive, our work is complementary to that of Ref.~\onlinecite{Kwasigroch2014} which considered the dynamics of non-equilibrium dipole-coupled two-level systems in a Ramsay sequence where the dynamics occurs in the absence of any drive field.

The paper is organised as follows. In Section~\ref{sec:setup} we define the model we study. We describe the resulting dynamical evolution in Section~\ref{sec:classicaldynamics} as a function of the strength of the Rabi coupling. We focus first on a  collective spin picture, \S\ref{subsec:collectivedynamics}, before proceeding to analyze when this collective picture breaks down, \S\ref{subsec:stability}. In \S\ref{subsec:analytical}, we study the high Rabi coupling limit analytically and show that in certain regimes collective Rabi oscillations are stable. The overall results are discussed in Section~\ref{sec:discussion} and the conclusions are summarized in Section~\ref{sec:conclusions}.

\section{Model}
\label{sec:setup}

\begin{figure}[H]
	\center
	\includegraphics[scale=0.4,clip,angle=0]{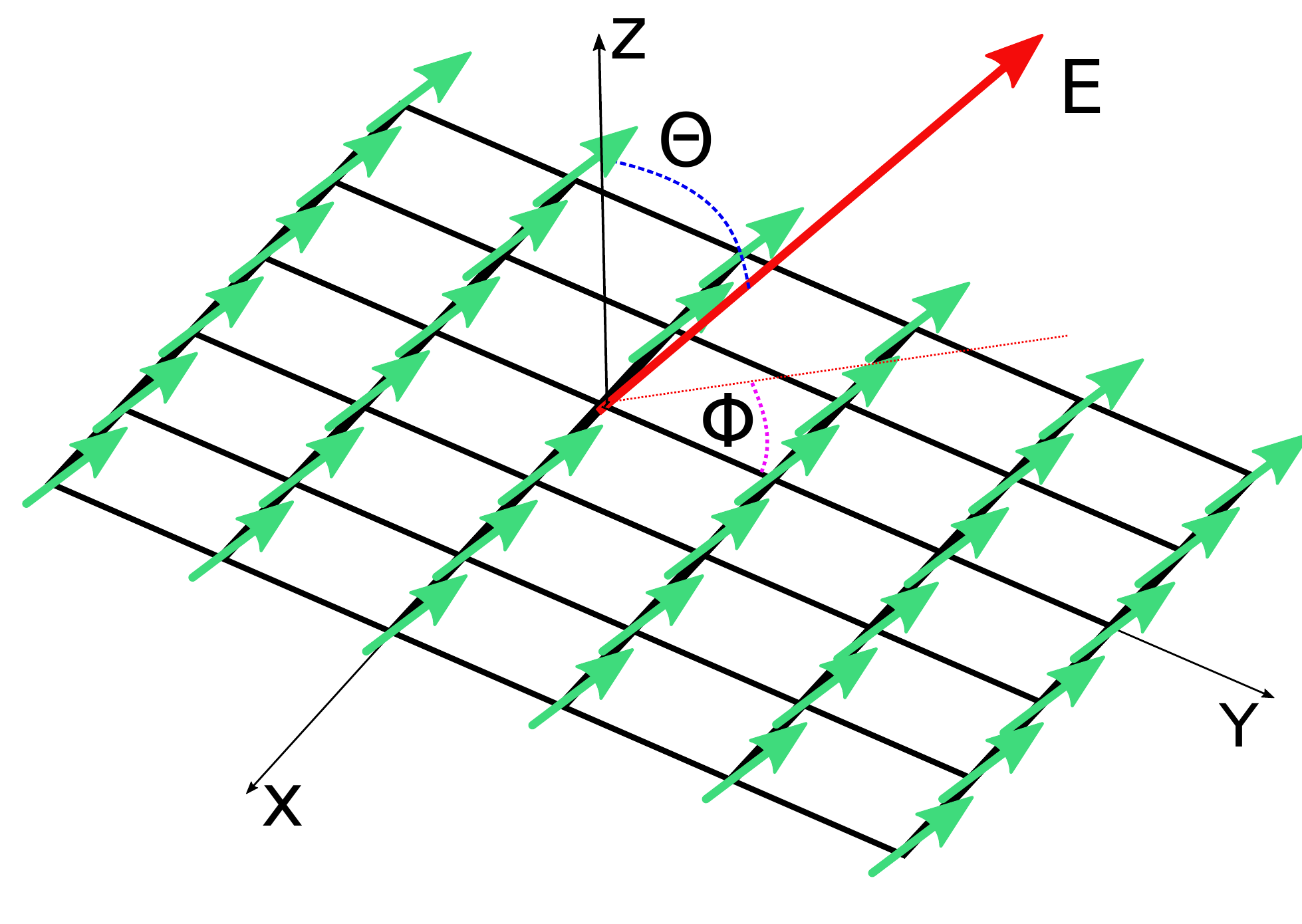}
	\vspace{-0cm}
	\caption{A sketch of the considered geometry, with dipoles oriented at an angle $\Theta$ to a 2D lattice under an applied external field, \textbf{E}. $\phi$ is the 2D polar angle between dipoles.}
	\label{Geometry}
\end{figure}

We study the dynamics of a large number of two-level systems coupled
by dipole interactions in a uniform external driving field.  Such a
system can be realised with atoms or polar molecules, held at fixed
positions in an optical lattice, with the two levels being two
internal energy levels between which there is a non-zero (magnetic or
electric) dipole matrix element (e.g. these could be two rotational
levels of a polar molecule).  Representing each two-level system by a
$S=1/2$ spin, and working in the rotating wave approximation such that 
the drive field is a 
time-independent Rabi coupling, leads to the Hamiltonian
\begin{equation}\label{HRWA}
\begin{split}
&\hat{H}=\Omega\sum_{i}^{N}\hat{S}_{i}^{x}+
\sum_{i\ne j}^{N}\frac{J_{ij}}{R_{ij}^3}\left[\cos\alpha \hat{S}_{i}^{z}\hat{S}_{j}^{z}+\sin\alpha\left(\hat{S}_{i}^{y}\hat{S}_{j}^{y}+\hat{S}_{i}^{x}\hat{S}_{j}^{x}\right)\right],
\end{split}
\end{equation} 
where $\Omega$ is the Rabi frequency. (We have set $\hbar=1$, which is used throughout this paper.) We shall consider the positions of the two-level systems to be in 1D or 2D arrays, as specified by the couplings $J_{ij}$. Here, $J_{ij}=J_0a^3(1-3\cos^2\phi_{ij}\sin^2\Theta)$ where $a$ is the lattice spacing, $\Theta$ represents the angle between the dipole orientation and the $z$ axis and $\phi_{ij}$ is the angle between dipoles in the $x-y$ plane (see Figure~\ref{Geometry}). For polar molecules, $\Theta$ can be tuned by application of a d.c. electric field along which the dipoles align \cite{Peter}. 
The parameter $\alpha$ relates the relative strength of the Ising and XY dipole interactions and can take values between $-\pi$ to $\pi$; we focus only on the range $0\leq \alpha\leq \pi$ as values below zero simply corresponds to a change in the sign of $J_0$ and give the same results. The value of $\alpha$ depends on the choice of internal states and external fields, with $\cos\alpha$ being related to the difference in dipole moments of the groundstate and excited state, and $\sin\alpha$ being related to the transition dipole moment between the groundstate and excited state. One such realisable set up involves the rotational states of polar molecules, the two levels being the 
 ro-vibrational groundstate ($J=0$) and one of the rotationally excited states ($J=1,m_J=0$).
However, other states could also be chosen \cite{Gorshkov2011}. The dynamics of the system with no driving has already been studied \cite{Peter2012,Model1982,Search}. The full model we consider has been touched upon in \cite{Schachenmayer2015}, but that work does not consider the driven steady states that we identify here. 
 Eq. \eqref{HRWA} can also be mapped to a hardcore boson model \cite{Zuniga}.

We assume that all the two-level systems are initially in the groundstate, corresponding to all the spins lying along $S_z=-1/2$. This initial state is not an eigenstate of Hamiltonian \eqref{HRWA} and so the subsequent evolution involves far-from-equilibrium dynamics.  Since we consider a macroscopic number of dipoles an exact description of the dynamics is impossible. We make the approximation that the system behaves classically and quantum fluctuations are negligible. We expect this approximation to be valid in regimes where the two-level systems remain largely coherent -- that is, with only long-range variations of the spin, such that there is a well-defined local spin and 
the system can be viewed as having slowly-varying ferromagnetic order. Indeed, we shall show that such situations of stable long-range coherence arise in certain parameter regimes.  

\section{Dynamical Evolution}
\label{sec:classicaldynamics}
In order to investigate the collective dynamics induced by Eq. \eqref{HRWA}, we solve numerically the Heisenberg equations of motion.
For an exact quantum system, the number of equations of motion scales as $2^N$ where $N$ is the number of spins, making the problem intractable for large systems. However, in the classical approximation, we can factorize $\langle\hat{S}^\alpha_i\hat{S}^\beta_j\rangle\approx\langle\hat{S}^\alpha_i\rangle\langle\hat{S}^\beta_j\rangle$ so quantum correlations are ignored. The number of equations of motion then scales linearly with system size and are given by

\begin{equation}\label{IndividualSpinEOMSHighFreq}
\begin{split}
&\frac{d S^{x}_i}{dt} = 2\sum_{j(\ne i)}^{N}\frac{J_{ij}}{R_{ij}^3}\left(\sin\alpha S^{z}_iS^{y}_j-\cos\alpha S^{z}_jS^{y}_i\right),\\
&\frac{d S^{y}_i}{dt} =-\Omega S^{z}_i-2\sum_{j(\ne i)}^{N}\frac{J_{ij}}{R_{ij}^3}\left(\sin\alpha S^{x}_iS^{z}_j-\cos\alpha S^{x}_jS^{z}_i\right),\\
&\frac{d S^{z}_i}{dt} =\Omega S^{y}_i+2\sin\alpha\sum_{j(\ne i)}^{N}\frac{J_{ij}}{R_{ij}^3}\left( S^{x}_iS^{y}_j- S^{x}_jS^{y}_i\right).\\
\end{split}
\end{equation} 
where $S_i^\alpha \equiv \langle \hat{S}_i^\alpha \rangle $.
We impose periodic boundary conditions, which in the large $N$ limit approximate a system with open boundary conditions very well for 1D and 2D, as the boundary effects become negligible. This would not be the case in 3D (which we do not consider here), for which the choice of boundary remains important for the dynamics owing to the dipolar interactions.

Solving Eqs. \eqref{IndividualSpinEOMSHighFreq} we find that the spins behave collectively if all prepared in the same initial state. We can therefore simplify the equations of motion and study the collective dynamics in more detail, before analysing the stability under perturbations.

\subsection{Collective Dynamics}
\label{subsec:collectivedynamics}
\begin{figure*}
	\center
	\includegraphics[scale=0.65,clip,angle=0]{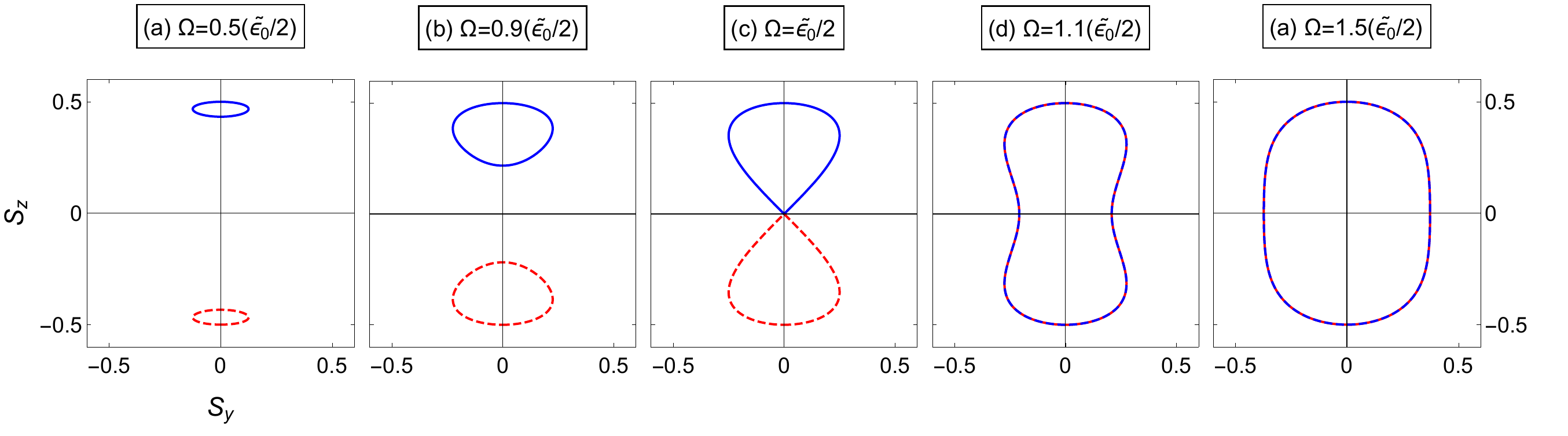}
	\vspace{-0cm}
	\captionsetup{skip=0pt}
	\caption{Bifurcation Dynamics
		(a) - Small oscillations at low Rabi coupling.
		(b) - Before the critical Rabi coupling at $\Omega=0.9(\tilde{\epsilon}_0/2)$. The two orbits for the spin up/down initial condition form teardrop shapes and start to meet at the origin which corresponds to the stationary state $(S_x,S_y,S_z)=(1/2,0,0)$.
		(c) - Orbit at the critical Rabi coupling, $\Omega=\tilde{\epsilon}_0/2$. The spin up and spin down orbits meet at the origin and form Homoclinic orbits. This change in dynamics is a Homoclinic Bifurcation.
		(d) - Past the critical Rabi coupling, $\Omega=1.1(\tilde{\epsilon}_0/2)$. The two orbits have now merged and explore each other's phase space. As the frequency increases, the orbit becomes a circle centred about the origin.
		(e) - Rabi Oscillations at high Rabi coupling.}
	\label{bifplots}
\end{figure*}

Treating the spins as a uniform collective spin, we set $S^{\alpha}_i=S^{\alpha} $ in Eqs. \eqref{IndividualSpinEOMSHighFreq}, and obtain a new set of equations valid on each lattice site
\begin{equation}\label{collectiveEquationsOfMotion}
\begin{split}
&\frac{d S^{x}}{dt} = 2\tilde{\epsilon_{0}}S^{z}S^{y},\\
&\frac{d S^{y}}{dt} =-\Omega S^{z}-2\tilde{\epsilon_{0}}S^{x}S^{z},\\
&\frac{d S^{z}}{dt} =\Omega S^{y},\\
\end{split}
\end{equation} 
with $\tilde{\epsilon_{0}}=(\sin\alpha-\cos\alpha)\sum_{\substack{ i\ne 0}}^{N}J_{i0}$ which is a measure of the dipole interaction strength, depending on the dipole tilt, underlying lattice and dimensionality. Figure \ref{bifplots} shows the spin dynamics for the specific case of $\alpha=\pi/2$ and with the dipoles perpendicular to the lattice ($\Theta=\pi/2$). We plot the orbit for an initial condition of $S_z=-1/2$  in red (dashed line), and that for an initial condition of $S_z=+1/2$ in blue (solid line). The equations of motion conserve total spin, so, for clarity, the  orbits can be projected down from the Bloch sphere onto the $S_z-S_y$ plane.

At low Rabi coupling, Figure~\ref{bifplots} (a), the orbits are small oscillations about the stationary states given by $(S_x,S_y,S_z)=(-\Omega/2\tilde{\epsilon}_0,0,\pm\frac{1}{2}\sqrt{1-\left(\Omega/\tilde{\epsilon}_0\right)^2}) $ 
As Rabi coupling increases, the orbit radius grows in size and the stationary state moves closer to the origin.
 At high Rabi coupling, Figure \ref{bifplots} (e), the orbits have merged to give full Rabi oscillations about the stationary state  $(S_x,S_y,S_z)=(1/2,0,0)$ 
 This behaviour can be derived from the collective equations of motion, Eqs. \eqref{collectiveEquationsOfMotion}, as $\tilde{\epsilon}_0$ becomes much smaller than $\Omega$ or zero in the special cases of dipole geometry or $\alpha=\pi/4$.

Figure \ref{bifplots} (b), (c) and (d) show the dynamics at intermediate Rabi couplings. At a critical coupling, $\Omega=\tilde{\epsilon_0}/2$, there is a bifurcation in the dynamics, indicative of a sharp transition in the form of the expected time evolution of the system. We can study this bifurcation more clearly by writing $S_x= \sqrt{S^2 - S_y^2+S_z^2}$ and defining new variables  $x=\sqrt{2\tilde{\epsilon_0}}S_z/\Omega$ and $y=\sqrt{2\tilde{\epsilon_0}}S_y$. This allows us to rewrite the $S_z$ and $S_y$ equations of motion, Eqs. \eqref{collectiveEquationsOfMotion}, as
\begin{equation}\label{DUFFING}
\begin{split}
& \frac{dx}{dt} =y\\ 
& \frac{dy}{dt} = -\Omega^2\left(1-\frac{\tilde{\epsilon_0}^2}{2\Omega^2}\right)x-\Omega^2\tilde{\epsilon_0}x^3
\end{split}
\end{equation}
The Hamiltonian of the system can also be recast as
\begin{equation}\label{DuffingHamiltonian}
H = \frac{1}{2}\Omega^2\left(1-\frac{\tilde{\epsilon_0}^2}{2\Omega^2}\right)x^2+\frac{1}{4}\Omega^2\tilde{\epsilon_0}x^4+\frac{1}{2}y^2.
\end{equation}
The collective equations of motion are now seen to be the equations of motion of a particle in a quartic potential well, which is also an undamped, unforced Duffing oscillator \cite{Author}. The potential energy forms a double well, with the spin initial conditions $S^z=\pm1/2$ sitting in one well each. At the bifurcation point, the two orbits hit the top of their respective potential wells and then move beyond them. Therefore, the Homoclinic orbit that arises at the critical frequency is also a separatrix between two different forms of orbits, those inside the double potential well and those outside. We can calculate the value of Rabi coupling at the bifurcation point by equating the energy of the orbit with that of the stationary state $(S_x,S_y,S_z)=(1/2,0,0)$, which gives the bifurcation coupling of $\Omega=\tilde{\epsilon_0}/2$ found in the dynamics.

In summary, within the assumption of a uniform collective spin, the dynamics undergo a transtion from small constrained orbits dominated by the dipole interaction at low Rabi coupling to full Rabi oscillations at large Rabi coupling. The boundary between these two regimes is given by a bifurcation in the dynamics.


\subsection{Stability to Non-Uniform Modulations}
\label{subsec:stability}
Having established the collective behaviour of the system, it is important to analyse the stability of the collective motion under small perturbations. To quantify this, we linearise the individual spin equations of motion, Eqs. \eqref{IndividualSpinEOMSHighFreq}, about the collective motion by setting $\textbf{S}_i=\textbf{S}_{col}+\delta \textbf{S}_i$ where $\textbf{S}_i$ is the spin vector on lattice site $i$. We then use the periodicity of the system to Fourier transform the linearised equations and put into matrix form $d\delta \textbf{S}_{\textbf{k}}/dt =\textbf{A}_{\textbf{k}} (t)\delta \textbf{S}_{\textbf{k}}$ where the matrix $\textbf{A}_{\textbf{k}}(t)$ contains the collective solutions and is therefore periodic. This means that our matrix equation is now a Floquet equation. We can write the solutions to this equation as a matrix that obeys $d\textbf{X}(t)/dt=\textbf{A}_{\textbf{k}}(t)\textbf{X}(t)$ which allows us to define the Monodromy matrix $\textbf{B}=\textbf{X}(T)$. The Monodromy matrix propagates the solutions forward by one period such that $\textbf{X}(t+T)=\textbf{B}\textbf{X}(t)$. 
After $N$ periods, we have $\textbf{X}(t+NT)=\textbf{B}^N\textbf{X}(t)$ and so any individual eigenstate of $\textbf{B}$ obeys $\textbf{x}_l(t+NT)=\rho_l^N\textbf{x}_l(t)$, with $\rho_l$ the eigenvalue, which we refer to as a Floquet multiplier. We calculate the full spectrum of Floquet multipliers and plot the largest as a function of Rabi coupling, where any multiplier with magnitude greater than unity represents an instability in the orbit as it leads to exponential growth of fluctuations.

In our analysis of the dynamics of the collective spin of \S\ref{subsec:collectivedynamics}, the tilt of the dipoles, $\alpha$, acted only to rescale the parameter $\tilde{\epsilon_0}$ and therefore we considered $\Theta=\pi/2$ as a representative case. However, in analysing stability to non-uniform spin-states, the tilt of dipoles can have significant effects. We therefore study three distinct geometries which are representative of all cases. The first is where all the dipoles are perpendicular to the lattice, $\Theta=0$, giving an isotropic environment at each site. The second is where the dipoles lie in the lattice with $\Theta=\pi/2$. Finally, we consider the dipoles at the so called `Magic Angle' for which the average dipole interaction vanishes. For 1D, this angle occurs at $\sin^2\Theta=1/3$ and for 2D occurs at $\sin^2\Theta=2/3$. We analyse the stability for each of the three dipole arrangements for specific cases of $0\leq \alpha\leq \pi$.

\subsubsection{Isotropic}
\begin{figure*}[h]
	\center
	\includegraphics[scale=0.44,clip,angle=0]{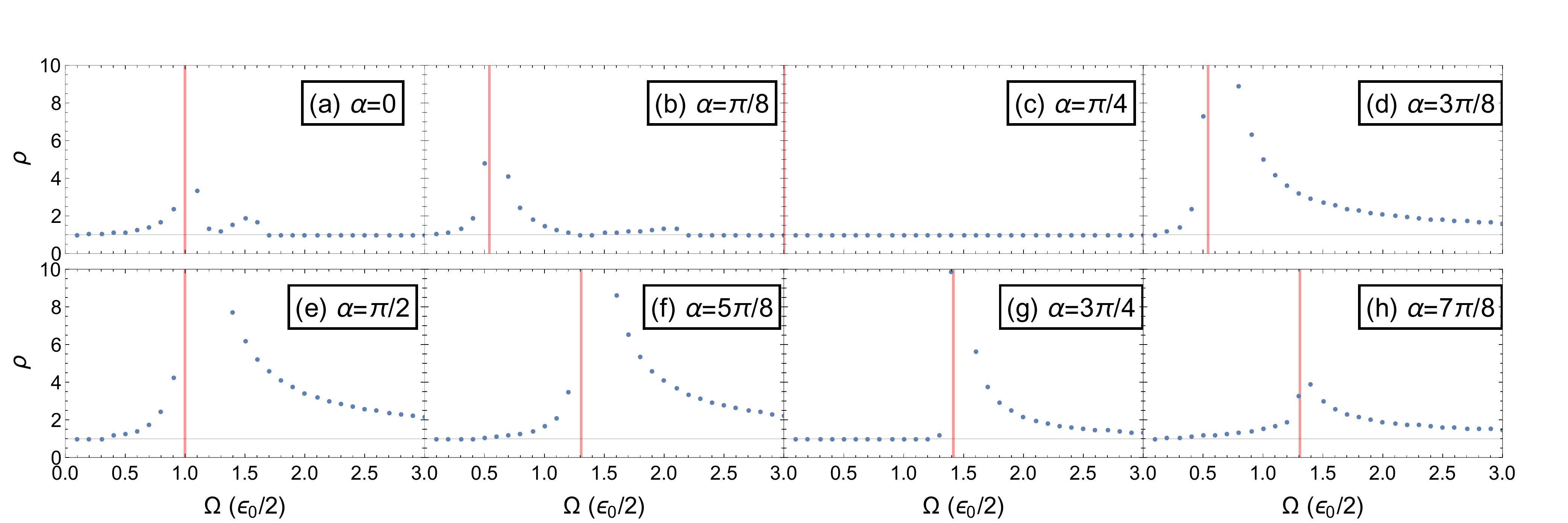}
	\vspace{-0cm}
	\captionsetup{skip=0pt}
	\caption{Plots of the largest Floquet multiplier for the 1D system. The multiplier is plotted as a function of Rabi coupling. The bifurcation line is shown in red. For each plot, the system is unstable for low Rabi coupling across the bifurcation point. However, for $\alpha<\pi/4$, the multipliers are in general smaller than for $\alpha>\pi/4$ and also the curves return to unity at high Rabi coupling.}
	\vspace{-2cm}
	\label{Isotropic Floquet}
\end{figure*}

The first case we consider is the isotropic dipole arrangement. Figure \ref{Isotropic Floquet} shows the largest Floquet multiplier for specific values of $\alpha$ between $0$ and $\pi$ in the 1D system. We find that for all values of $\alpha$, the initial instability behaviour is the same. At low coupling, the multiplier is near unity, but increases in size as a function of Rabi coupling. Stability is expected at zero coupling given that the initial state is an eigenstate of the Hamiltonian. As the Rabi coupling approaches the bifurcation value, the collective dynamics becomes increasingly unstable, reaching maximum values at the bifurcation point.  Increasing the Rabi coupling further, past the bifurcation point, the multiplier decreases in value indicating a suppression of the instability. Therefore, we find that, in general, the collective picture is unstable and the only places where it may persist are high Rabi couplings. Our numerics show that in the high Rabi coupling region, stability depends on the value of $\alpha$. For $\pi/4<\alpha\leq7\pi/8$, our numerics show the Floquet multiplier never reaches unity except for $\alpha=3\pi/4$, meaning the collective Rabi oscillations always break apart. However, for $0\leq\alpha\leq\pi/4$, the multiplier tends to unity again and collective motion persists. Later, we will quantify analytically the high Rabi coupling behaviour and derive the regime of stability.

Moving away from the high Rabi coupling region, we find that for $0\leq\alpha\leq\pi/4$, the multiplier is close to unity for a Rabi coupling that is comparable to, but still greater than the dipole interaction. We also note the presence of oscillations in the multiplier value which are robust to changes in system size and geometry. These oscillations and the stability down to relatively low values of the 
Rabi coupling stability are unexplained, but do not affect our conclusions in the high Rabi coupling regime.

Across the $\alpha$ range, the bifurcation point is seen to move closer to zero Rabi coupling as we move $\alpha$ from $0$ through to $\pi/4$. At $\alpha = \pi/4$ there is complete stability for all values of the Rabi coupling. This can be understood as a consequence of the conservation of total spin which occurs as the Hamiltonian becomes an isotropic Heisenberg Hamiltonian. 

\begin{figure*}
	\center
	\includegraphics[scale=0.44,clip,angle=0]{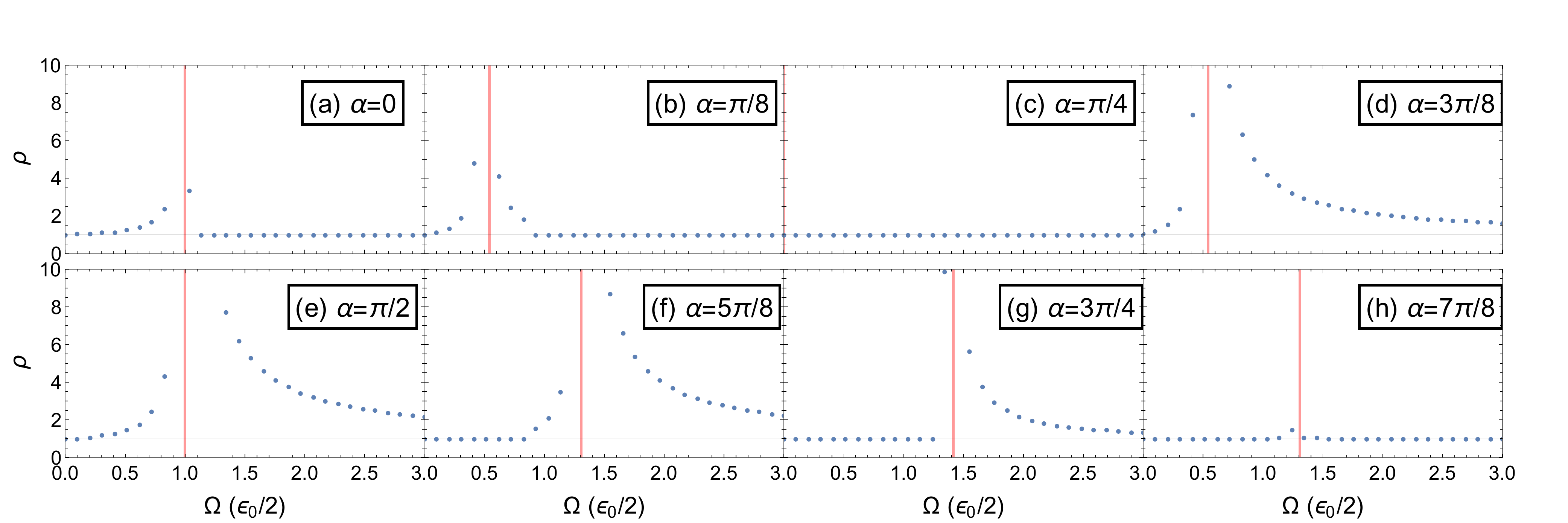}
	\vspace{-0cm}
	\captionsetup{skip=0pt}
	\caption{Plots of the largest Floquet multiplier for the 2D isotropic system, with the bifurcation line shown in red. The multiplier is plotted as a function of Rabi coupling. The results are similar to those in 1D, but with stability arising for $\alpha=7\pi/8$ at high Rabi coupling.}
	\vspace{-2cm}
	\label{2D Isotropic Floquet}
\end{figure*}

In Figure \ref{2D Isotropic Floquet}, we plot the largest Floquet multiplier for the 2D system for the same $\alpha$ values. The qualitative results are the same as the 1D system, with the only difference being stability for the case $\alpha=7\pi/8$ in 2D at high Rabi coupling. 

Overall, in both 1D and 2D, we find stability of the collective state can only be obtained either at high Rabi coupling, zero Rabi coupling or when $\alpha=\pi/4$. Small perturbations otherwise broaden the bifurcation peak in the collective dynamics.

\begin{figure*}[h]
	\center
	\includegraphics[scale=0.44,clip,angle=0]{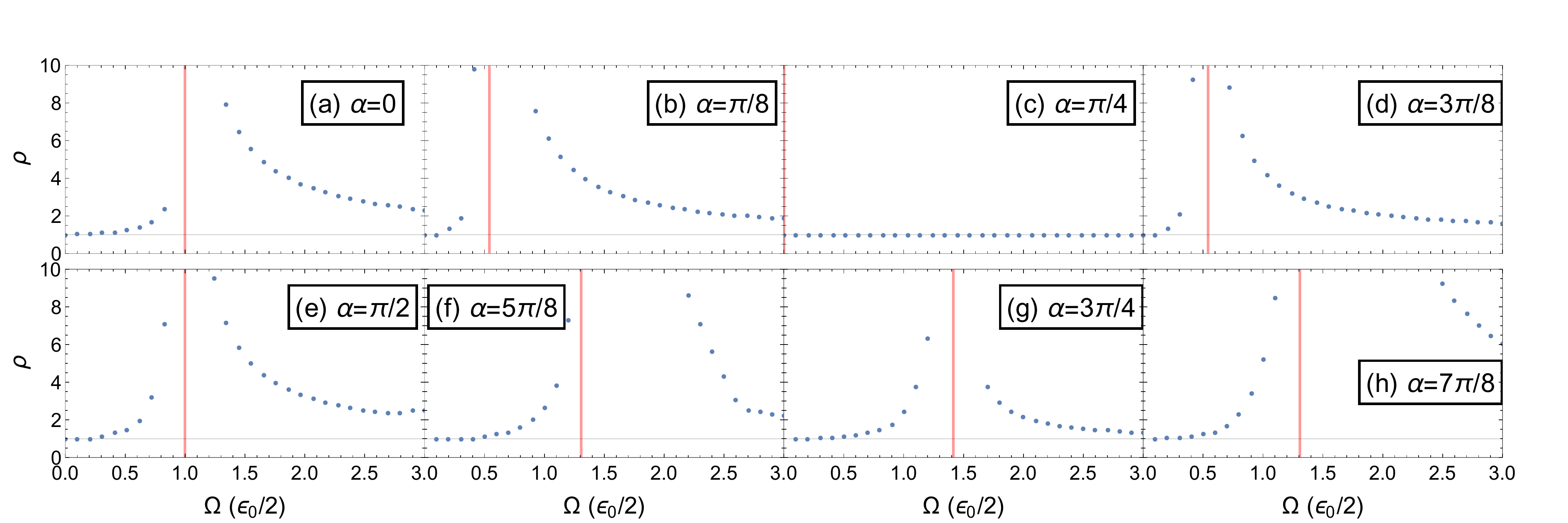}
	\vspace{-0cm}
	\captionsetup{skip=0pt}
	\caption{Plots of the largest Floquet multiplier for a 2D lattice for an in plane geometry. The multiplier is plotted as a function of Rabi coupling. The bifurcation line is shown in red. For each plot, we see the system is unstable for low Rabi coupling across the bifurcation point, but also unstable at high Rabi couplings in contrast to the isotropic case.}
	\vspace{0cm}
	\label{In Plane Floquet}
\end{figure*}

\subsubsection{In Plane}
We now consider the case where the dipoles lie along the lattice, $\Theta = \pi/2$. For the 1D case, there is no change in stability due to the fact that tilting the dipoles only changes the relative sign of the interaction. Therefore, the results of the stability analysis will mirror that for the 1D isotropic case. For 2D however, the tilting of dipoles creates anisotropy in the dipole interaction and will have different effects. Figure \ref{In Plane Floquet} shows the Floquet multiplier results. The same broad instability behaviour about the bifurcation coupling is seen, but in contrast to the isotropic case, the only situation where stable orbits occur at high Rabi coupling is when $\alpha=\pi/4$.

\subsubsection{Magic Angle}
\begin{figure*}
	\center
	\includegraphics[scale=0.44,clip,angle=0]{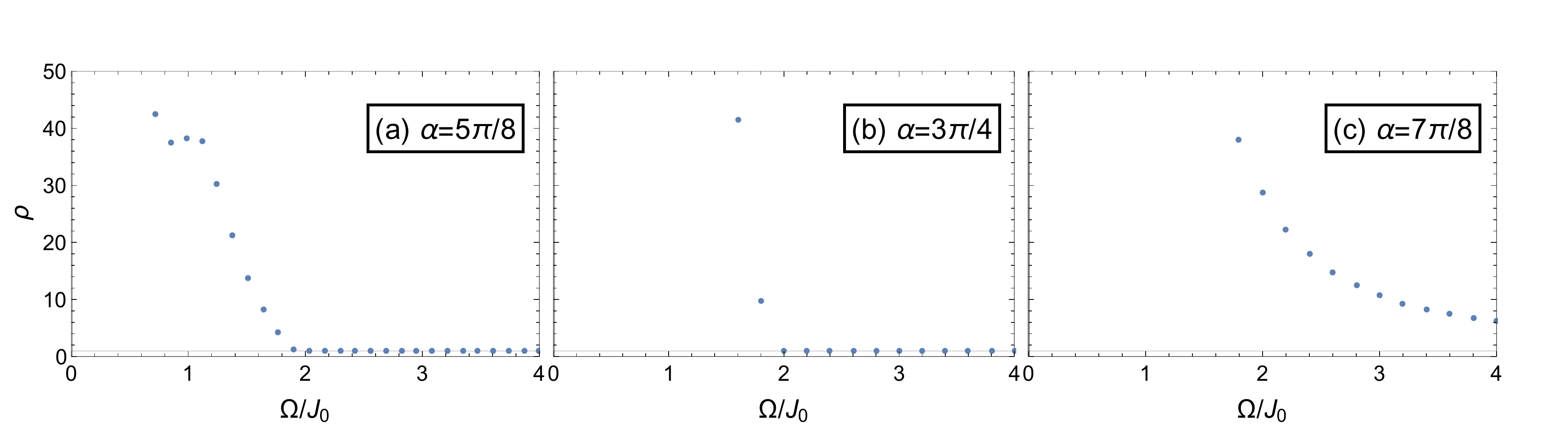}
	\vspace{-0cm}
	\captionsetup{skip=0pt}
	\caption{Plots of the largest Floquet multiplier for a 2D system at the magic angle, $\sin^2\Theta = 2/3$. The multiplier is plotted as a function of Rabi coupling. The system is stable at high Rabi coupling for $0\leq \alpha \leq 3\pi/4$, and unstable for $3\pi/4<\alpha<\pi$.}
	\label{Critical Tilt Floquet}
\end{figure*}

Finally, we consider the case where the dipoles are oriented at the magic angle which leads to an average zero dipole interaction. 
For 1D, the dipole interaction cancels entirely and so stable Rabi oscillations are seen across the entire Rabi coupling range for all $\alpha$ values. For 2D, the dipole interaction only cancels on average and so instability can still arise from fluctuations. We find that at low Rabi coupling, fluctuations give rise to large instabilities, but at large Rabi coupling, the system becomes stable again for $0\le\alpha\le3\pi/4$ and remains unstable for $3\pi/4<\alpha<\pi$. Because of the average dipole cancellation, the time period of the collective state depends purely on the Rabi coupling and there is no longer a bifurcation or critical coupling. In Figure \ref{Critical Tilt Floquet} we  show the Floquet multiplier results for $\alpha=5\pi/8$, $\alpha=3\pi/4$ and $\alpha=7\pi/8$. Any oscillatory behaviour in the Floquet multiplier at low coupling is a finite size effect. 

\subsection{Analytical Analysis at High Rabi Coupling}
\label{subsec:analytical}
Our numerical Floquet analysis showed that collective behaviour was only stable in the high Rabi coupling limit under certain ranges of $\alpha$ and with certain tilt angles of the dipoles. Here, we analytically compute when the collective picture is stable in the high Rabi coupling limit and compare to our numerical results. To do this, we move to a frame rotating at the Rabi frequency, obtaining a time-dependent Hamiltonian with no Rabi term to which we apply a Magnus expansion, as outlined in Ref.~\cite{Goldman2015}, in powers of $J_0/\Omega$. The full derivation can be found in Appendix A.

To zeroth order in the expansion, the effective high frequency Hamiltonian is given as
\begin{equation}\label{Heff}
\hat{H}_{\rm eff} = \sum_{i\ne j}^{N}\frac{J_{ij}}{R_{ij}^3}\left[\sin \alpha \hat{S}_{i}^{x}\hat{S}_{j}^{x}+ \frac{\sin\alpha+\cos\alpha}{2} \left(\hat{S}_{i}^{y}\hat{S}_{j}^{y} + \hat{S}_{i}^{z}\hat{S}_{j}^{z}\right)\right].
\end{equation}
We now study the stability of persistent Rabi oscillations by performing a spin wave analysis of the effective Hamiltonian about the $S_z$ collective state. The resulting Hamiltonian can be diagonalized to find the normal-mode frequencies, leading to the spin-wave excitation spectrum

\begin{equation}\label{Excitation Spectrum}
 E_\textbf{q} =\frac{1}{2} \sqrt{(\epsilon_\textbf{q}-\epsilon_0)(\sin\alpha+\cos\alpha)((2\epsilon_\textbf{q}-\epsilon_0)\sin\alpha-\epsilon_0\cos\alpha)},
\end{equation}
with
\begin{equation}\label{Dipole Dispersion}
\begin{split}
& \epsilon_{\textbf{q}}= \sum_{i}^{N}J_0a^3(1-3\cos^2\phi_{\textbf{R}_i}\sin^2\Theta)\frac{e^{i\textbf{q}\cdot\textbf{R}_i}}{|\textbf{R}_i|^3},
\end{split}
\end{equation}
where $\phi_{\textbf{R}}$ is the angle between the position vector and the $x$ axis. By studying when the excitation spectrum becomes imaginary, we can find when Rabi oscillations are unstable. Also note that as the momentum tends to zero, the excitation spectrum also tends to zero. This is consistent with the presence of a Goldstone boson, related to the continuous symmetry of the Hamiltonian under rotations
around $S_x$.

Before examining the excitation spectrum for each of the three geometries, we establish the low momentum limit of the dipole dispersion, Eq. \eqref{Dipole Dispersion}, which is important in determining the stability of Rabi oscillations. In 1D, $\phi_\textbf{R}=0$ and so we can expand Eq. \eqref{Dipole Dispersion} directly to obtain the following low momentum behaviour 
\begin{equation}\label{1D low momentum}
\epsilon_{q} = \left(1-3\sin^2\Theta\right)\left(\epsilon^{\rm (1D)}_0-a^2q^2+O(q^4)\right),
\end{equation}
where $\epsilon^{\rm (1D)}_0 \approx 2.4 J_0$.
In 2D, due to the angle $\phi_\textbf{R}$, we need to separate the dipole dispersion into different angular components
\begin{equation}\label{Dipole Dispersion separated}
\begin{split}
& \epsilon_{\textbf{q}}=\epsilon_{\textbf{q}}^0\left(1-\frac{3}{2}\sin^2\Theta\right)-\frac{3}{2}\sin^2\Theta\left(\epsilon_{\textbf{q}}^{(2)}+\epsilon_{\textbf{q}}^{(-2)}\right),
\end{split}
\end{equation}
where
\begin{equation}\label{Angular}
 \epsilon_{\textbf{q}}^{(m)}=  J_0a^3\sum_{i}^{N}\frac{e^{i\textbf{q}\cdot\textbf{R}_i+im\phi_{\textbf{R}}}}{|\textbf{R}_i|^3},
\end{equation}
For a large number of spins, the spectrum is conveniently found via Ewald summation \cite{Peter,Bonsall1977}. This involves rewriting the dispersion in terms of Bessel functions of the first kind, which on expansion give
\begin{equation}\label{2D low momentum}
\begin{split}
&\epsilon_{\textbf{q}}=\epsilon^{\rm (2D)}_0\left(1-\frac{3}{2}\sin^2\Theta\right)\\
&+2|\textbf{q}|\pi\left(-1+\frac{3}{2}\sin^2\Theta-\sin^2\Theta\cos2\psi\right)\\
&+|\textbf{q}|^2\left(1-\frac{3}{2}\sin^2\Theta-\eta\sin^2\Theta\cos2\psi\right).
\end{split}
\end{equation}
where $\eta\simeq 0.23$ is a numerical constant. Here, $\epsilon^{\rm (2D)}_0 \approx 9 J_0$ and $\psi$ is the angle between the $\textbf{q}$ vector and the $q_x$ axis. The key difference between 1D and 2D is the presence of the linear term in momentum. Given the low momentum expansion of the dipole interactions, we can now study the stability of Rabi oscillations for the different geometries.


\subsubsection{Isotropic}
For the isotropic case, stability is predominantly determined by low momentum fluctuations. Applying the low momentum expansions in the excitation spectrum gives $ E_\textbf{q} = \sqrt{\epsilon^{\rm (2D)}_0\cos(2 \alpha)(2\pi |\textbf{q}|a)}$ in 2D and  $E_\textbf{q} = \sqrt{\epsilon^{\rm (1D)}_0\cos(2 \alpha)(2\pi q^2a^2)}$ for 1D.
We find that stability depends only on the value of $\alpha$.
For $0 \leq\alpha \leq \pi/4$, the spectrum is real and therefore we have stable Rabi oscillations. However, for $\pi/4 < \alpha < 3\pi/4 $, the spectrum becomes imaginary and Rabi oscillations become unstable. Both these outcomes agree with the numerical analysis. We can see here that $\alpha=\pi/4$ and $\alpha=3\pi/4$ result in zero excitation spectrum. When $3\pi/4<\alpha$, the spectrum becomes imaginary as high momentum fluctuations cause instabilities instead of low momentum ones,  but this is the only region where this occurs. The spectrum is real again when $7\pi/8<\alpha$ in 1D and $2.506<\alpha$ in 2D. 

We can go further than simply the prediction of instability to compare our analytical results to our numerical results and can calculate the decay rate of the collective state in the unstable regions. From the spectrum of the $S^z$ state, the time scale of decay is derived from the momentum that causes maximum instability i.e. gives the largest imaginary component. When $\pi/4<\alpha<3\pi/4$, this maximum decay occurs when $\epsilon_\textbf{q}=(3\sin\alpha+\cos\alpha)\epsilon_0/4\sin\alpha$, which gives an energy and corresponding decay time of
\begin{equation}\label{Emax}
\text{Im}(E)=1/\tau=\frac{\epsilon_0(\sin\alpha-\cos\alpha)}{4\sqrt{2}}\sqrt{\frac{\sin\alpha+\cos\alpha}{\sin\alpha}},
\end{equation}
where we have assumed the expectation value of the spins decays as $\exp(iEt)$ and once again set $\hbar=1$.
However, when $3\pi/4<\alpha<\pi$, there are still decay modes in certain cases, and these occur for $q=\pi$. 
The Floquet multipliers relate to the decay of the initial state in the following way
\begin{equation}\label{NT}
\textbf{x}(t+NT)=\rho^N\textbf{x}(t),
\end{equation}
where $\textbf{x}(t)$ is a vector of the classical solutions and the Floquet multiplier sets the growth rate of fluctuations/decay of the collective orbit. Given an initial state $\textbf{x}(t)$, we can relate these two time scales by
\begin{equation}\label{DecayTimeCalc}
\log\left(\frac{\textbf{x}(t+NT)}{\textbf{x}(t)}\right)=N\log(\rho)=\frac{NT}{\tau},
\end{equation}
where $\tau=1/\text{Im}(E)$. Therefore,
\begin{equation}\label{DecayTimeTwo}
\tau=\Lim{\Omega \rightarrow \infty}\frac{T}{\log(\rho)}.
\end{equation}

\begin{figure}[H]
	\center
	\includegraphics[scale=0.45,clip,angle=0]{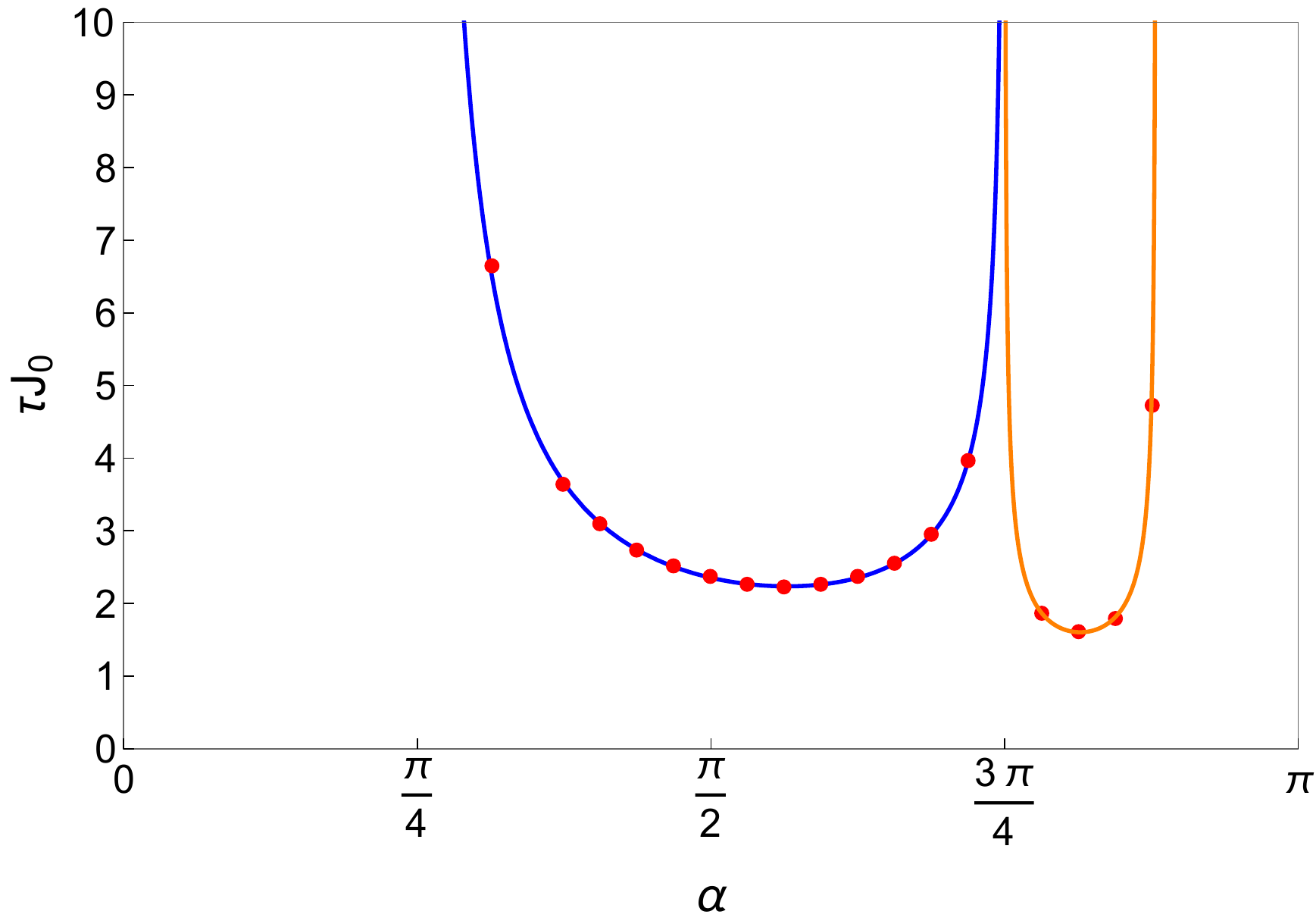}
	\vspace{-0cm}
	\caption{Decay Times for the 1D Floquet system. The Blue (darker) curve is the analytic curve for decay from Eq. \eqref{Emax}, the orange (lighter) curve is the excitation spectrum evaluated at $q=\pi$ and the points are the Floquet multipliers for $\Omega=50\epsilon_0$.  The numerically computed Floquet multipliers are very close to the analytic curve.}
	\label{DecayTImes}
\end{figure}

Figure \ref{DecayTImes} shows the decay times for different values of $\alpha$ for the 1D isotropic case plotted with the theoretical curves from our analytic results. The blue (darker) curve is the decay time from \eqref{Emax} and the orange (lighter) curve from \eqref{Excitation Spectrum} at $q=\pi$. The agreement is excellent.

\subsubsection{In Plane}
As we mentioned before in the Floquet analysis, \S\ref{subsec:stability}, there is no change in the stability of the 1D dipole system when the dipoles lie in the lattice plane as the head-head interactions simply becomes a head-tail interaction and leads to a change in sign of $J_0$. We can see this directly from the low momentum expansion Eq. \eqref{1D low momentum}. However, in 2D, there is now both head-head and head-tail interactions which gives an anisotropy to the dipole interaction. The 2D low momentum expansion, Eq. \eqref{2D low momentum}, has a negative quadratic dispersion along one direction, and positive linear dispersion along another due to the $\cos2\psi$ term. This means that the excitation spectrum Eq. \eqref{Excitation Spectrum} is unstable at both low and high momenta for all values of $\alpha$ as along certain directions, the spectrum is imaginary. This again agrees with our numerical results which showed instability at high Rabi coupling for all values of $\alpha$ except $\alpha=\pi/4$ and $\alpha=3\pi/4$.

\begin{figure}[H]
	\center
	\includegraphics[scale=0.45,clip,angle=0]{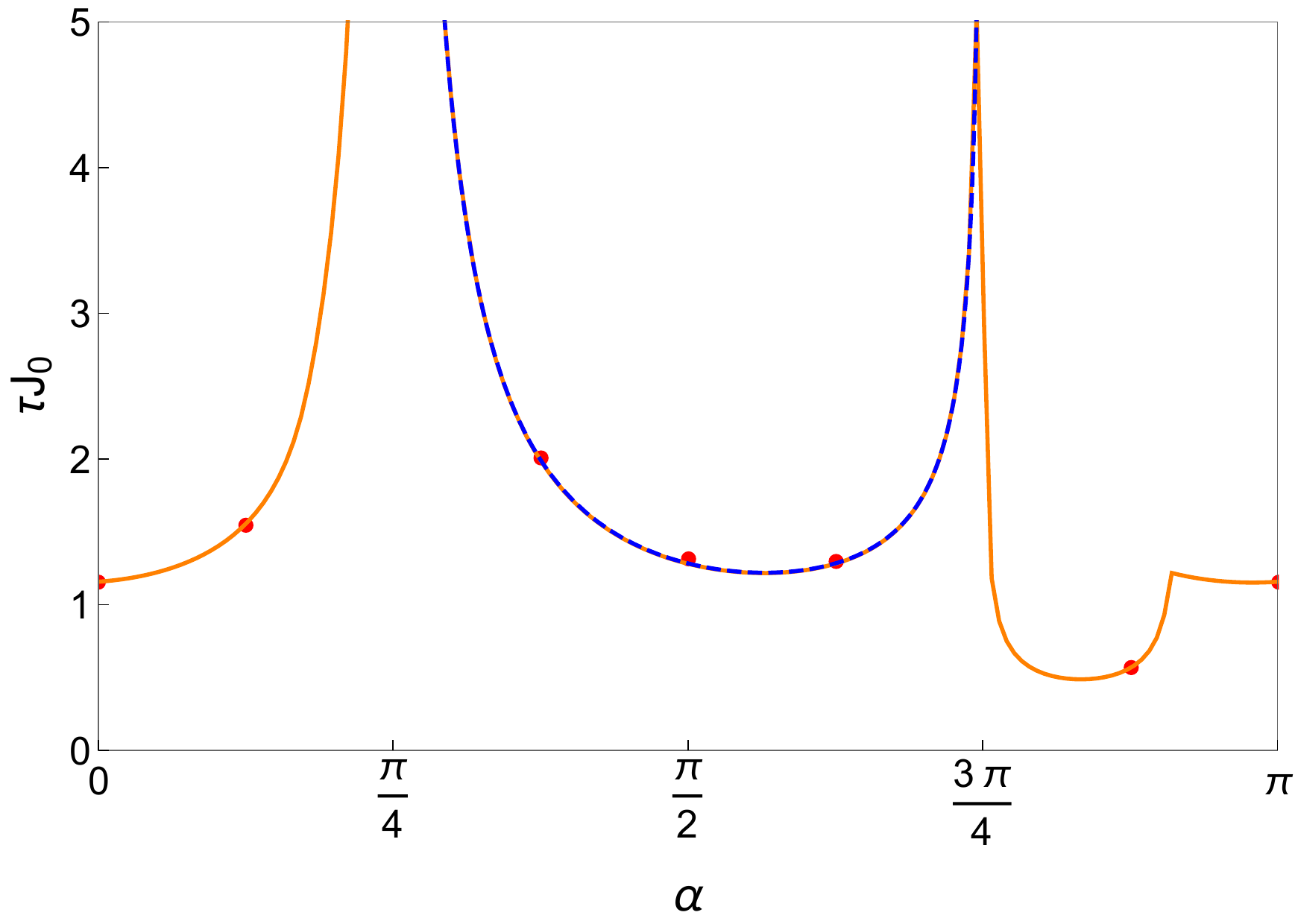}
	\vspace{-0cm}
	\caption{Decay Times for the 2D system with dipoles in plane. The blue (darker) curve is the analytic curve from the $\alpha$ decay and the orange (lighter) curve from numerical analysis of Eq. \eqref{Excitation Spectrum}. The points are the Floquet multipliers evaluated at $\Omega=50\epsilon_0$. The numerically computed Floquet multipliers match the analytic curve very well.}
	\label{2DIPDecayTimes}
\end{figure}

As for the isotropic case, we can compute decay times to compare analytics and numerics. These are plotted in figure  \ref{2DIPDecayTimes} for the 2D system. Although not plotted here the 1D curve is identical to that from the isotropic case. We see that the 2D data fits the theoretical curve well. 

\subsubsection{Magic Angle}
Lastly, we consider the case where the dipoles lie at the magic angle. For 1D, the dipole interaction disappears completely, so $\epsilon_q=0$ for all momenta and there are no effects of spin-wave fluctuations at any Rabi coupling. For 2D, whilst $\epsilon_0=0$, the dipole interaction only cancels along certain directions. The fluctuation spectrum becomes
\begin{equation}\label{Tilt Dipole Interaction}
E_\textbf{q} = \epsilon_\textbf{q}\sqrt{2\sin\alpha(\sin\alpha+\cos\alpha)}.
\end{equation}
The average cancellation of the dipole interaction allows us to pull the dipole dispersion out of the square root. Now the stability of this spectrum depends purely on the relative strengths of the Ising and XY terms in the Hamiltonian and becomes unstable if $3\pi/4<\alpha<\pi$, as seen in our numerical results.

\subsubsection{Higher Order Corrections}
In our spin wave analysis, we have only focused on the leading term in the Magnus expansion, finding it gives excellent agreement with our numerical results at high Rabi coupling. We expect terms of order $1/\Omega$ (see Eq.\eqref{FirstOrderCorrectionApp} in Appendix A) or higher in the Magnus expansion will also capture lower Rabi coupling behaviour under spin wave analysis, but have the additional complication of the kick operator. Study of the effective Hamiltonian by itself at higher orders results in a conservation of total $S_x$ which is not physical at lower Rabi coupling and so the kick operator (which violates total $S_x$ conservation) must be included to capture the full system dynamics. However, this has no influence on our results, which are at zeroth order, as the leading term of the kick operator is of order $1/\Omega$.

\section{Discussion}
\label{sec:discussion}
\begin{figure}
	\center
	\includegraphics[scale=0.6,clip,angle=0]{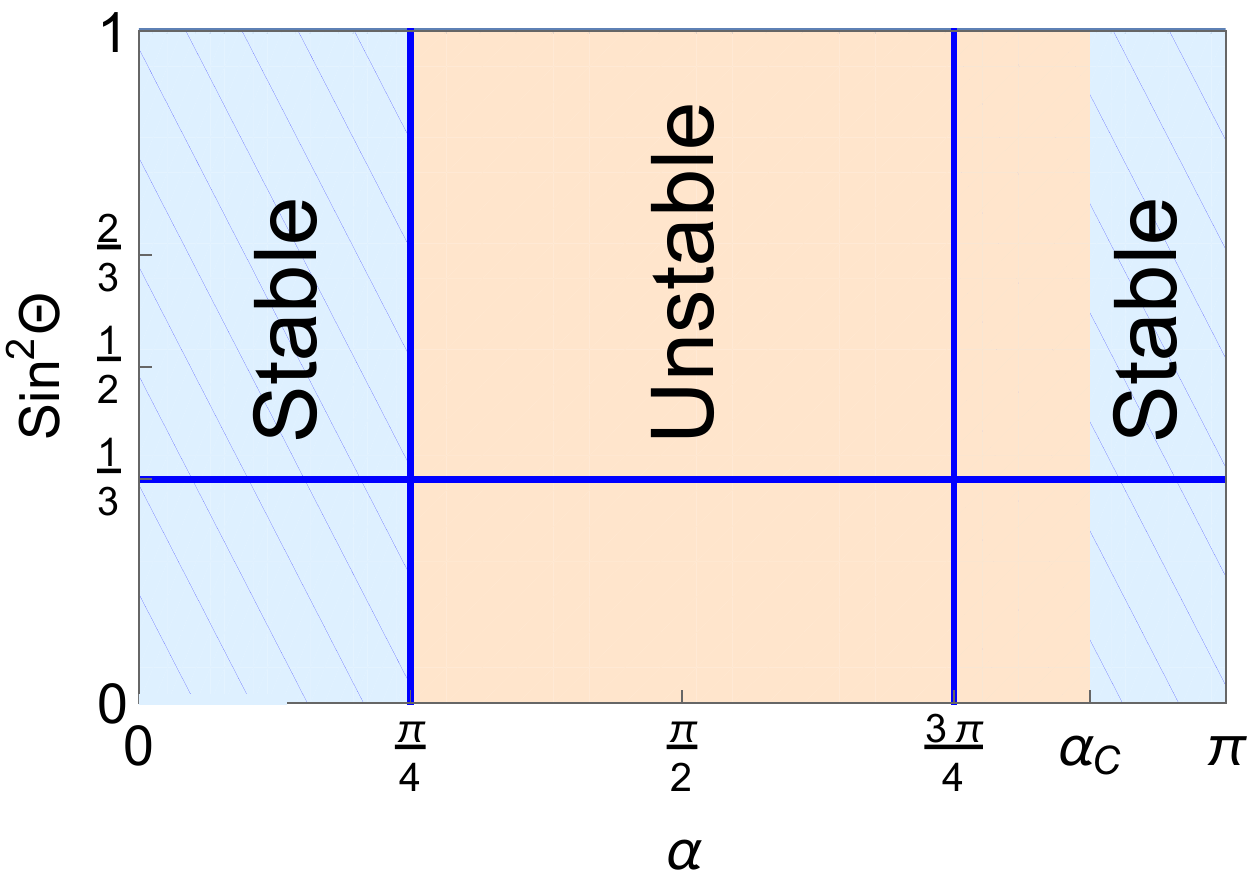}
	\vspace{-0cm}
	\caption{Phase Diagram at high Rabi coupling in 1D. The blue (lined) regions represent stability and the orange (plain) regions represent instability. The $\alpha=\pi/4$ line is stable across all tilt angles due to the Heisenberg symmetry. The magic angle line at $\sin^2\Theta=1/3$ is stable across the entire $\alpha$ range and occurs due to the cancellation of the entire dipole interaction. There is a region of instability between $3\pi/4<\alpha\leq7\pi/8$ due to high momentum fluctuations.}
	\label{PhaseDiagramTilt1D}
\end{figure}

\begin{figure}
	\center
	\includegraphics[scale=0.6,clip,angle=0]{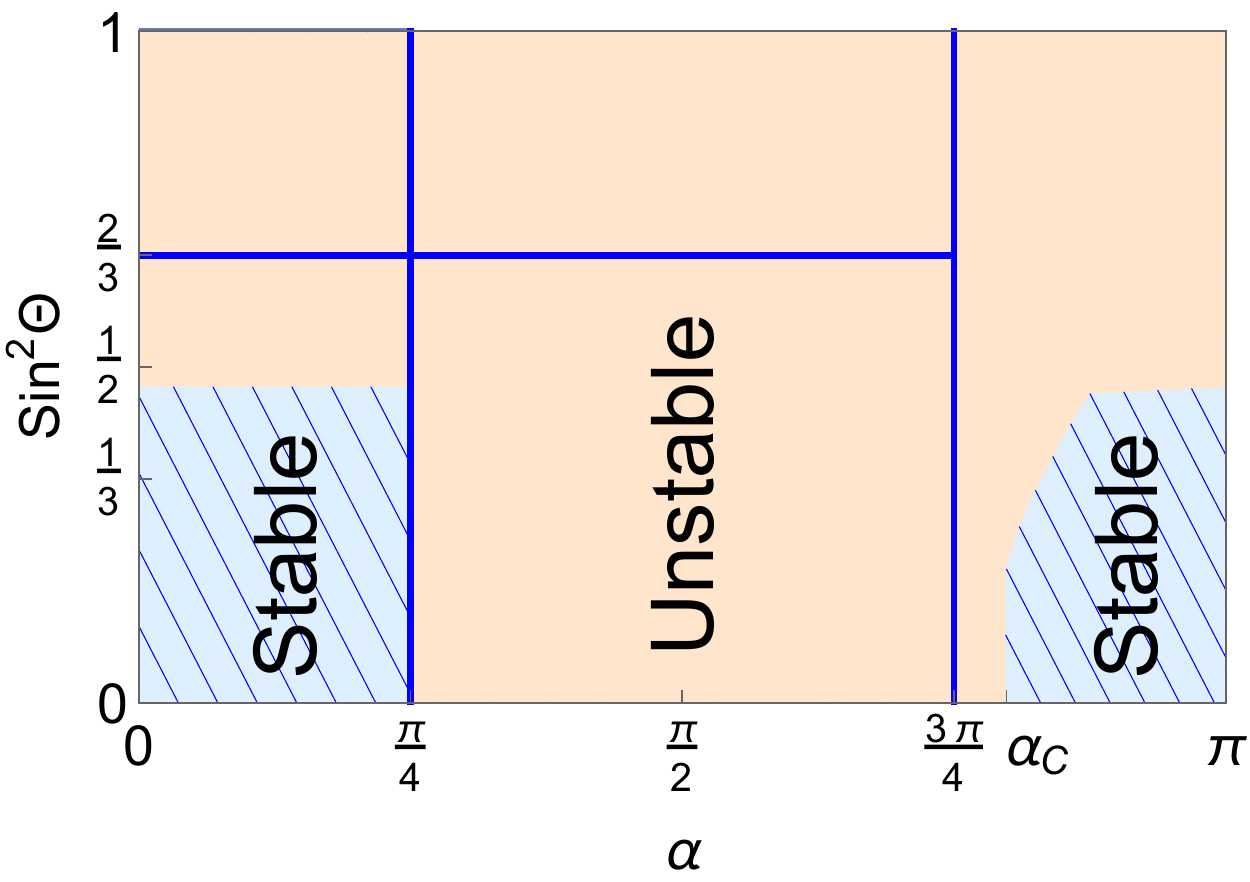}
	\vspace{-0cm}
	\caption{Phase Diagram at high Rabi coupling in 2D. The blue (lined) regions represent stability and the orange (plain) regions represent instability. As for 1D, the $\alpha=\pi/4$ line is stable across all tilt angles due to the Heisenberg symmetry. The magic angle line at $\sin^2\Theta=2/3$ is stable up to $\alpha=3\pi/4$. Unlike in 1D, the anisotropy in the dipole interaction causes instability as the dipoles are tilted for $0\leq\alpha\leq\pi/4$.}
	\label{PhaseDiagramTilt}
\end{figure}

We have established the form of the collective dynamics of the system and the regimes and geometries under which the coherent collective dynamics of all two-level systems is stable against non-uniform modulations. We find in both 1D and 2D that stability occurs across the entire Rabi coupling range for $\alpha=\pi/4$ and otherwise stability occurs only at high Rabi coupling under certain conditions and dimensionality. In this high Rabi coupling limit, one can determine stability of the collective picture from the spectrum of spin waves of the effective Hamiltonian, Eq. \eqref{Excitation Spectrum}. We use this to plot a phase diagram of the system in the high Rabi coupling regime as a function of tilt angle $\Theta$ and the relative strengths of the Ising and XY interaction, $\alpha$. These phase diagrams are shown in Figures \ref{PhaseDiagramTilt1D} and \ref{PhaseDiagramTilt} for the 1D and 2D system respectively. 

Summarising the key features of the phase diagrams, we find that for  the isotropic geometry, $\sin^2\Theta=0$, only the regions $0\leq\alpha\leq\pi/4$, $\alpha=3\pi/4$ or $\alpha_c<\alpha\leq\pi$ leads to stable Rabi oscillations in 1D and 2D. For $\pi/4<\alpha<3\pi/4$, instabilities arise from low momentum fluctuations whilst for $3\pi/4<\alpha<\alpha_c$, instabilities arise from high momentum fluctuations. The cut off angle $\alpha_c$ is the value of $\alpha$ beyond which high momentum fluctuations cannot cause instability and we find these values take $\alpha^{\rm (1D)}_c=7\pi/8$ and $\alpha^{\rm (2D)}_c=2.506$ for 1D and 2D respectively. Increasing the tilt of the dipoles, we find these conditions of stability do not change in 1D where tilting the dipoles only scales the interaction strength, except at the magic angle where the dipole interactions cancel exactly. However, in 2D, the collective picture becomes unstable for all $\alpha$ past a certain value of $\sin^2\Theta$, when the anisotropy in the dipole dispersion becomes large enough to destabilise the excitation spectrum. We also find that $\alpha^{\rm (2D)}_c$ depends on the tilt angle. Once the tilt reaches the magic angle, the average dipole interaction cancels and we again obtain stability across all values of $\alpha$ except in the range $3\pi/4<\alpha<\pi$ for 2D.

The existence of stable collective Rabi oscillations over a range of Rabi couplings at the magic angle is surprising in 2D, given that small scale fluctuations still exist and otherwise destabilise the collective picture at tilt angles above and below the critical line. It would be interesting to see how this depends on the underlying geometry and why the fluctuations can behave in such a way that stability is retained. In all other cases, the collective picture breaks apart after a finite time set by the value of the Floquet multiplier, with quickest decay of the collective dynamics occurring at the bifurcation coupling. By studying the time evolution of the system in the unstable regions via numerical simulations, we find that chaotic dynamics dominate and possible thermal states form. For example, at low Rabi coupling when $0<\alpha<\pi/4$, we find there is an average steady state of spins lying in the $xz$ plane. This result is similar to those of exact diagonalization results in a system with van der Waals interactions \cite{Lesanovsky2010}.

Our results in this section rely on two important approximations. One approximation is the truncation of the Magnus expansion to lowest order. This neglects terms of order $J_0^2/\Omega$, so our results should hold at least up to time scales of $t\approx \Omega/J_0^2$. While our results could persist much longer than this, the full Magnus expansion is expected to break down on  time scales that are exponentially long in $\Omega/J_0$, which sets an upper limit for the lifetime of stable oscillations before the system becomes ergodic and heats to infinite temperature \cite{Kuwahara2016}. The other approximation is to treat the effective Hamiltonian Eq. \eqref{Heff} within the classical approximation. Exact diagonalization of the effective Hamiltonian  shows that for small systems (up to $N=12$), the classical approximation breaks down due to spin squeezing \cite{Kitagawa1993}, where quantum fluctuations of the collective spin grow in the $zy$ plane. Consequently the classical approximation is only valid up to times of order $t\approx2\sqrt{N}/\tilde{\epsilon_0}$. Thus, for any finite systems the Rabi oscillations will dephase for all $\alpha$.

Finally, we make effort to relate our results to experiment. In our analysis, we have implemented periodic boundary conditions as a convenient way of studying the thermodynamic limit. From numerical simulations of the dynamics of systems with open boundary conditions in 1D and 2D, we find that the collective oscillation can break down more easily for small systems with open boundary conditions than for periodic boundary conditions. However, these finite size effects become insignificant at the high Rabi couplings for which we predict collective behaviour and our analytic analysis becomes valid. Indeed, the analytic results at high Rabi coupling are readily performed also for open boundary conditions, and lead to the same conclusions and same phase diagrams. Therefore, we expect our results to hold in systems that are large, but still well below the thermodynamic limit.

Furthermore, in Figures \ref{DecayTImes} and \ref{2DIPDecayTimes}, we have calculated the dephasing time of our system which we now compare to other experimental time scales. If we consider a system of KRb molecules with $J_0/h = 52 \text{Hz}$, this gives a decay time of around $2s$ (for $\alpha=\pi/2$). This is much lower than the lifetime of the molecules (up to $25s$) \cite{Hazzard2014} so the breakdown of Rabi oscillations should be evident in experiment. Similarly, for Rydberg atoms under resonant exchange ($\alpha=\pi/2$), we find dephasing time scales of $25\mu s$ and Rydberg lifetimes of up to $100\mu s$ in the absence of motional dephasing \cite{Maxwell2013}, so again, we expect our results to be observable.

Lastly, we studied briefly numerics of a non-unity filling fraction and detuning which can arise in experimental set up. We find that at high Rabi coupling, any effects from filling disorder appeared to not be significant for $0\leq\alpha<\alpha_c$. Inhomogeneity did seem to cause decay for $\alpha_c<\alpha<\pi$ where before the system was stable, but with large decay times. Therefore, experimentally, the Rabi oscillations will eventually breakdown in this region. For detuning disorder, we found no change in stability of Rabi oscillations, which agrees with fact detuning appears as a $1/\Omega$ term in our Magnus expansion. Data for both disorder types is shown in the Appendix B.

\section{Conclusions} 
\label{sec:conclusions}
We have studied a system of interacting dipoles with both an Ising and XY dipole interaction terms under external driving in both 1D and 2D. After preparing all the spins in an ideal $S_z$ state, we found that collective behaviour occurs, with Rabi oscillations at high Rabi coupling, small oscillations at low Rabi coupling and a bifurcation in the dynamics at a critical value of coupling. Using Floquet analysis, we find that the collective picture is unstable for most values of Rabi coupling. However, we find stable collective behaviour at high coupling under certain regimes, either when the dipoles are oriented at the magic angle or when the Ising interaction dominates over the XY interaction. 
We have identified regimes of parameters where the driven dynamics of the two-level systems shows stable collective coherent Rabi oscillations despite the presence of the dipole-dipole coupling.

\section{Acknowledgements}
C. P. would like to thank A. Lamacraft for helpful discussions and comments. This work was supported by EPSRC Grant No. EP/K030094/1. Statement of compliance with EPSRC policy framework on research data: All data accompanying this publication are directly available within the publication.

\begin{appendices}
\section*{A - Magnus Expansion}
Below is a derivation of the effective Hamiltonian given by Eq. \eqref{Heff} and the first order correction.
We start with the time dependent Hamiltonian we obtain from applying the unitary transform $\hat{U}=\exp[i\Omega\sum_{i}^{N}\hat{S}^x_i]$ to Eq. \eqref{HRWA}. Splitting into time independent and dependent parts gives us
\begin{equation}\label{Split}
\begin{split}
&\hat{H}_{rot} = \hat{H}_{0}+\hat{V}(t)=\sum_{i\ne j}^{N}\frac{J_{ij}}{R_{ij}^3}\left(\hat{S}_{i}^{x}\hat{S}_{j}^{x}+\frac{1}{2} \hat{S}_{i}^{y}\hat{S}_{j}^{y}+\frac{1}{2} \hat{S}_{i}^{z}\hat{S}_{j}^{z}\right)\\
& +\sum_{i\ne j}^{N}\frac{J_{ij}}{R_{ij}^3}\left(\frac{\cos(2\Omega t)}{2} (\hat{S}_{i}^{y}\hat{S}_{j}^{y}-\hat{S}_{i}^{z}\hat{S}_{j}^{z})-\sin(2\Omega t) \hat{S}_{i}^{y}\hat{S}_{j}^{y}\right) \\
& +\sum_{i}^{N}\Delta_{i}\left(\hat{S}_{i}^{z}\cos(\Omega t)+\hat{S}_{i}^{y}\sin(\Omega t)\right).
\end{split}
\end{equation}
where $\Delta_{i}\equiv \omega_0-\omega$ is the detuning of the applied field frequency $\omega$ from the two level transition frequency $\omega_0$. We now write the potential in terms of Fourier components
\begin{equation}
\begin{split}
& \hat{V}^{\pm 1}=\sum_{i}^{}\frac{\Delta_i}{2}\left(\hat{S}_{i}^{z}\mp i\hat{S}_{i}^{y}\right),\\
& \hat{V}^{\pm 2}=\sum_{i\ne j}^{}\frac{J_{ij}}{4 R_{ij}^3}\left[\hat{S}_i^y \hat{S}_j^y-\hat{S}_i^z \hat{S}_j^z\pm i\hat{S}_i^y \hat{S}_j^z\right].
\end{split}
\end{equation}
Following \cite{Goldman2015}, we can now make the following unitary transform, $\hat{U}=\exp[i\hat{K}(t)]$, to move to a rotating frame that removes the time dependent parts. This gives the effective Hamiltonian 
\begin{equation}
\begin{split}
\hat{H}_{\rm eff} = e^{i\hat{K}(t)}\hat{H}(t)e^{-i\hat{K}(t)}+i\frac{\partial e^{i\hat{K}(t)}}{\partial t}e^{-i\hat{K}(t)}.
\end{split}
\end{equation}
We can expand $\hat{K}$ and $\hat{H}$ in powers of $1/\Omega$ and by comparing powers, obtain the following
\begin{equation}
\begin{split}
& \hat{H}_0 = \hat{H}_0,\\
& \hat{H}_{1}=\frac{1}{\Omega}\sum_{j=1}^{\infty}\frac{1}{j}\left[\hat{V}^{(j)},\hat{V}^{(-j)}\right],\\
& \hat{H}_{2}=\frac{1}{2\Omega^2}\sum_{j=1}^{\infty}\frac{1}{j^2}\left(\left[\left[\hat{V}^{(j)},\hat{H}_0\right],\hat{V}^{(-j)}\right]+h.c.\right),
\end{split}
\end{equation}
where $h.c.$ denotes the Hermitian Conjugate. The effective Hamiltonian is just given by $\hat{H}_0$. Given our above Fourier decomposition, we find the first term in the expansion as
\begin{equation}\label{FirstOrderCorrectionApp}
\frac{1}{2\Omega}\sum_{i}^{N}\Delta_{i}^2\hat{S}_{i}^{x} +\frac{\cos\alpha-\sin\alpha}{4\Omega}\sum_{\substack{i,k,l \\ i\ne l \\k\ne l}}^{N}\frac{J_{il}J_{kl}}{R_{il}^3R_{kl}^3}\left(\hat{S}_{i}^{y}\hat{S}_{k}^{y}+\hat{S}_{i}^{z}\hat{S}_{k}^{z}\right)\hat{S}_{l}^{x}.
\end{equation}
The Kick operator is given by
\begin{equation}\label{KickOp}
\begin{split}
&\hat{K}(t) =\frac{1}{2\Omega}\sum_{i}^{N}\Delta_{i}\left(\hat{S}_{i}^{z}\sin(2\Omega t)+\hat{S}_{i}^{y}\cos(2\Omega t)\right)+\\
&\frac{1}{\Omega}\sum_{i,j}^{N}\frac{J_{ij}}{R_{ij}^3}\left(\frac{\sin(2\Omega t)}{2}(\hat{S}_{i}^{y}\hat{S}_{j}^{y}-\hat{S}_{i}^{z}\hat{S}_{j}^{z})-\hat{S}_{i}^{y}\hat{S}_{j}^{z}\cos(2\Omega t)\right)
\end{split}
\end{equation}

\section*{B - Disorder Results}
Below we include data for both filling and detuning disorder. We calculate the Floquet multiplier and subsequent decay times at a Rabi coupling of $\Omega=50\epsilon_0$, which is well within the high Rabi coupling range. We look to see if disorder causes further instabilities than those already found in the main text.

\begin{figure}[H]
	\center
	\includegraphics[scale=0.4,clip,angle=0]{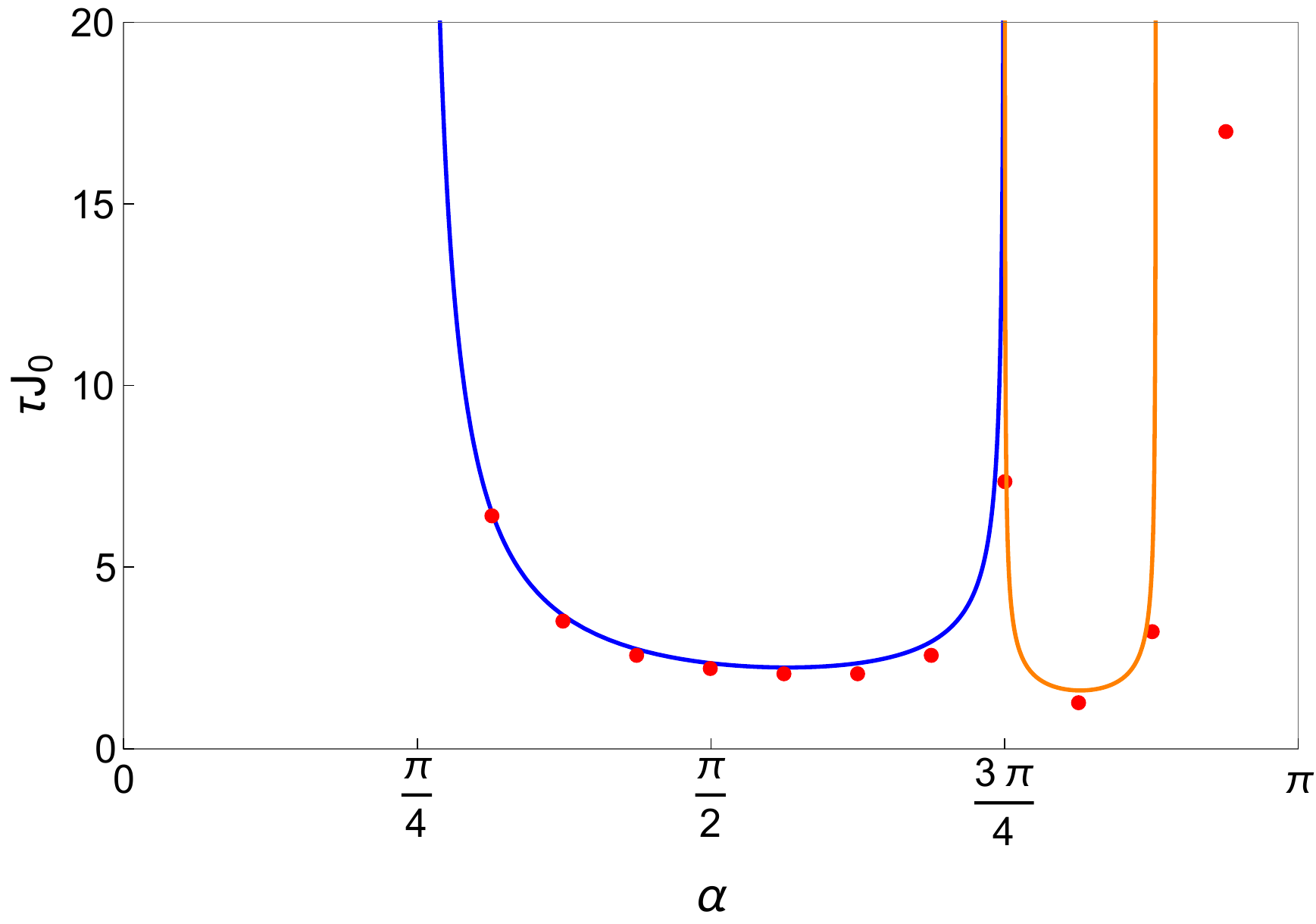}
	\vspace{-0cm}
	\caption{Decay times of a 1D system where each site has an occupation probability of $0.75$. We find that filling disorder can cause decay in the region  $\alpha_c<\alpha<\pi$ which is stable at unity filling, but with long decay times. Otherwise, filling disorder causes no other instabilities.}
	\label{fillingdisorder}
\end{figure}

We first consider non-unity filling disorder, where each site is randomly occupied or unoccupied with an occupation probability of $0.75$. The decay time is calculated for each configuration and averaged over 100 different configurations. Results are shown in Figure \ref{fillingdisorder}.

We find that the same stability/instability transition as for the homogeneous system, except in the range $\alpha_c<\alpha<\pi$ which becomes unstable, but with large decay times. Therefore, in this range, we expect breakdown of the Rabi oscillations will occur, but otherwise our results should hold in the presence of filling disorder.

\begin{figure}[H]
	\center
	\includegraphics[scale=0.4,clip,angle=0]{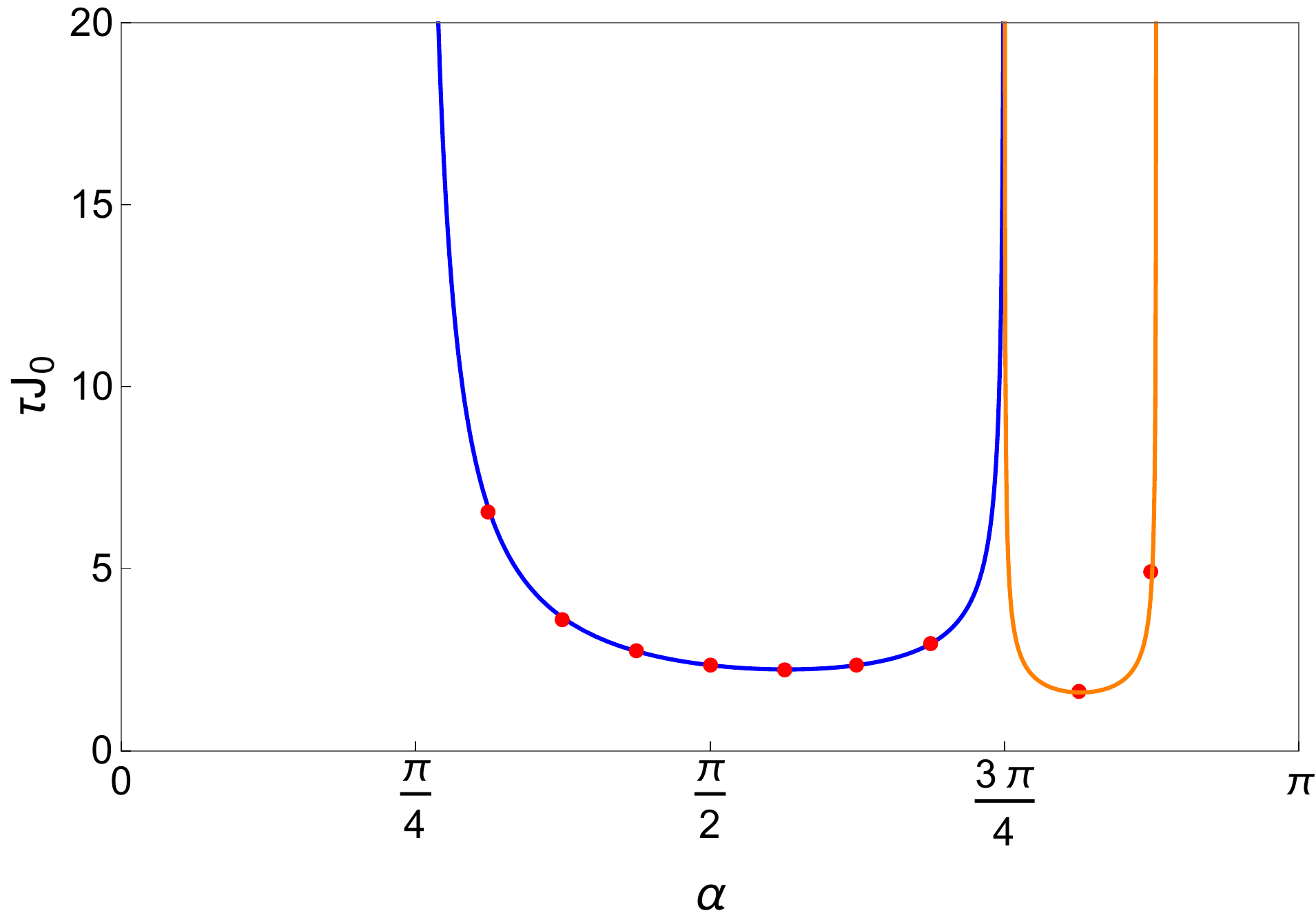}
	\vspace{-0cm}
	\caption{Decay times of a 1D system with small random detunings that are Gaussian distributed about zero with a standard deviation of $W=0.1J_0$. We find the detuning disorder does not cause further instabilities compared to the ordered system.}
	\label{detuningdisorder}
\end{figure}

We also calculate the decay time for small random detunings that are Gaussian distributed about zero with a standard deviation $W=0.1J_0$, and once again average over 100 configurations. The decay times are shown in Figure \ref{detuningdisorder}.
We find no change in stability across the entire range of $\alpha$. This agrees with our Magnus expansion, where detuning appears as a $1/\Omega$ term and so should be negligible at high enough Rabi coupling. Non-unity filling, however, appears even at zeroth order in the Magnus expansion as it modifies the dipole coupling $J_{ij}$. 
In both disorder studies, we have looked at a low level of disorder. We expect that at strong enough disorder, more instabilities will arise and Rabi oscillations will break down for all $\alpha.$

\section*{C - Floquet Theory}
Below we prove the key results from Floquet theory and how they relate to the stability of the system.
For any Floquet equation, we can write the solutions in matrix form
\begin{equation}\label{FloquetFormApp}
\frac{d\textbf{X}}{dt}=\textbf{A}(t)\textbf{X}.
\end{equation}
For any non-singular matrix $\textbf{X}(t)$, we can find another matrix $\textbf{Y}(t)=\textbf{X}(t)\textbf{B}$ that is also a solution to the equations where $B$ is some constant matrix
\begin{equation}
\begin{split}
\textbf{Y}'(t)&=\textbf{X}'(t)\textbf{B}\\
&=\textbf{A}(t)\textbf{X}(t)\textbf{B}\\
&=\textbf{A}(t)\textbf{Y}(t).
\end{split}
\end{equation}
Given this, we can show that $\textbf{X}(t+T)$ is a solution to Eq. \eqref{FloquetFormApp} provided $\textbf{A}(t+T)=\textbf{A}(t)$
\begin{equation}
\begin{split}
\textbf{Y}'(t)& =\textbf{X}'(t+T)\textbf{B} \\
& =\textbf{A}(t+T)\textbf{X}(t+T)\textbf{B}\\
&=\textbf{A}(t)\textbf{X}(t+T)\textbf{B}\\
&=\textbf{A}(t)\textbf{Y}(t).
\end{split}
\end{equation}
Given $\textbf{X}(t+T)=\textbf{X}(t)\textbf{B}$ is a solution and because $\textbf{B}$ is time independent, we can find $\textbf{B}$ for $t=0$ to give
\begin{equation}
\textbf{B}=\textbf{X}(0)^{-1}\textbf{X}(T).
\end{equation}
This is the Monodromy matrix whose eigenvalues give the Floquet multipliers. Each multiplier can be written as the exponent of a Floquet value. However, these values are not unique due to the addition of $2\pi i$ in the exponent.

We can show the multipliers relate to the stability of the orbits as follow. Let $\textbf{b}$ be an eigenvector of $\textbf{B}$ and let $\textbf{x}(t)=\textbf{X}(t)\textbf{b}$. Then we have
\begin{equation}
\textbf{x}(t+T)=\textbf{X}(t+T)\textbf{b}=\textbf{X}(t)\textbf{B}\textbf{b}=\rho\textbf{x}(t).
\end{equation}
Therefore, after time $T$, the solution to the Floquet equation becomes scaled by the multiplier. If we let time run for $NT$ times, we find
\begin{equation}
\textbf{x}(t+NT)=\rho^N\textbf{x}(t).
\end{equation}
We can see then that if the multiplier is greater than one, the solution will diverge exponentially in time. If the multiplier is unity, the solution is periodic and if the multiplier is less than unity, the solution decays. i.e. the fluctuations die down. Therefore, the original solution is only stable if all multipliers are less than unity. 

\end{appendices}

\nocite{*}

%

\end{document}